\documentclass[twocolumn,aps,showpacs,prl,amsmath,amssymb,floatfix,superscriptaddress]{revtex4}
\usepackage{graphicx}
\usepackage{dcolumn}
\usepackage{bm}
\usepackage{color} 
\usepackage{array}
\usepackage{float}
\usepackage{supertabular}
\usepackage{longtable}
\begin{document}

\newcommand{\naoki}[1]{\textcolor{blue}{\bf [Naoki: #1]}}

\title{Heterogeneous Voter Models}
\author{Naoki Masuda}
\affiliation{Graduate School of Information Science and Technology,
The University of Tokyo,
7-3-1 Hongo, Bunkyo, Tokyo 113-8656, Japan}
\affiliation{PRESTO, Japan Science and Technology Agency,
4-1-8 Honcho, Kawaguchi, Saitama 332-0012, Japan}
\author{N.~Gibert}
\affiliation{Ecole Nationale Sup\'erieure de Techniques Avanc\'ees, 32 Boulevard Victor, 75739 Paris, FRANCE}
\affiliation{Center for Polymer Studies and Department of Physics, Boston University, Boston, MA USA~ 02215}
\author{S.~Redner}
\affiliation{Center for Polymer Studies and Department of Physics, Boston University, Boston, MA USA~ 02215}

\begin{abstract}
  We introduce the {\em heterogeneous voter model\/} (HVM), in which each
  agent has its own intrinsic rate to change state, reflective of the
  heterogeneity of real people, and the {\em partisan voter model\/} (PVM),
  in which each agent has an innate and fixed preference for one of two
  possible opinion states.  For the HVM, the time until consensus is reached
  is much longer than in the classic voter model.  For the PVM in the
  mean-field limit, a population evolves to a preference-based state, where
  each agent tends to be aligned with its internal preference.  For finite
  populations, discrete fluctuations ultimately lead to consensus being
  reached in a time that scales exponentially with population size.
\end{abstract}


\maketitle

The paradigmatic voter model~\cite{L85} describes the evolution toward
consensus in a population of agents that possess a discrete set of opinions.
In a single update, a random voter is picked and it adopts the opinion state
of a randomly-selected neighbor.  By repeated updates, a finite and initially
diverse population necessarily reaches consensus in a time that typically
scales as a power law of the population size $N$~\cite{L85,K02}.  In many
respects, the voter model resembles the kinetic Ising model with
zero-temperature Glauber dynamics.  Because of this connection to
non-equilibrium spin systems~\cite{spin} and the utility of the voter model
for interacting particle~\cite{L85} and social~\cite{CFL} systems, the voter
model is widely-studied in the physics literature (see {\it e.g.},
\cite{BFK96,DCCH01,SR05,SAR08}).  In this work, we generalize the traditional
voter model in two simple, but far-reaching ways to incorporate the
heterogeneity of real people~\cite{FDC09}: 
\begin{itemize}
\item {\em Heterogeneous voter model} (HVM):  each voter has an intrinsic
  and distinct ``flip'' rate.
\item {\em Partisan voter model} (PVM): each voter has an innate and fixed
  preference for one opinion state.
\end{itemize}

The role of heterogeneity was emphasized in classic work by
Granovetter~\cite{G78}, in which collective social behavior is determined by
the diversity of individual thresholds to act in response to stimuli.  In the
context of the voter model, heterogeneity has been studied in the extreme
situation where some voters are ``zealots'' that never change
opinion~\cite{M03,GF07}.  This attribute prevents consensus from being
reached when zealots with different opinions exist.  In our HVM, the flip
rate of each agent is taken from a continuous distribution that excludes
zero.  Since every voter can, in principle, change state, a finite system
necessarily reaches consensus, albeit slowly.  For the PVM, the innate voting
preference of each agent leads to a collective state in which the opinion of
each voter tends to align with its own preference.  This competition between
self-interest and consensus has been modeled previously~\cite{G97}, and was
the focus of recent social experiments that attempted to elucidate the role
of the preference strength on the dynamics~\cite{KJTW}.  Here we investigate
basic properties of these two models from a statistical physics
perspective. \smallskip

\noindent{\em Heterogeneous Voter Model} (HVM): Each agent can be in one of
two opinion states that we label as $\mathbf{0}$ and $\mathbf{1}$.  We first
determine the exit probability $E(\rho)$ that a finite population with
initial density $\rho$ of $\mathbf{1}$ voters ends with $\mathbf{1}$
consensus.  Because the average density $\rho$ of $\mathbf{1}$ voters is
conserved for the classic voter model on regular networks~\cite{L85}, the
final density of $\mathbf{1}$ voters, which equals $0\times
(1-E(\rho))+1\times E(\rho)=E(\rho)$, must equal the initial density $\rho$.

To derive the exit probability for the HVM, we need to construct an analogous
conservation law.  Let $\eta(x)=0,1$ denote the state of a voter at node $x$
in a social network, $\eta$ the state of all voters in the system, and
$\eta^x$ the system state derived from $\eta$ when only the voter at $x$
flips.  The transition probability of a voter at node $x$ is given by
\begin{equation}
  \label{master}
  \textbf{P} [\eta \to \eta^{x}] = 
  \sum_{y} \frac{r_x}{Nk}\,[\Phi(x,y)+\Phi(y,x)],
\end{equation}
where $y$ are the neighbors of node $x$, $r_x$ is the intrinsic flip rate of
the voter at $x$, and $k$ is the number of neighbors of each node in a
regular network. The factor $\Phi(x,y)\equiv \eta(x)[1-\eta(y)]$ guarantees
that voters at $x$ and $y$ have different opinions so that an update actually
occurs.  The transition probability \eqref{master} corresponds to the {\em
  invasion process} on a heterogeneous network (see Eqs.~(4) and (5)
in~\cite{SAR08}), in which a randomly-selected agent imposes its state on a
neighbor; in the complementary voter model the agent imports the state of a
neighbor.

The average change in $\eta(x)$ equals the difference between the
probabilities that $\eta(x)$ changes from 0 to 1 and from 1 to 0.  Thus
$\langle\Delta\eta(x)\rangle = \left[1-2\eta(x)\right]
\mathbf{P}[\eta\rightarrow\eta_x]$.  Using the transition rate
\eqref{master}, it is immediate to see that the factor $r_x$ leads to
$\langle\eta(x)\rangle$ not being conserved.  By construction, however, the
rate-weighted density of $\mathbf{1}$ voters,
\begin{equation}
  \omega\equiv\frac{ \sum_x \eta(x)/r_x}{\sum_x 1/r_x}~,
\label{eq:omega}
\end{equation}
{\em is} conserved in the HVM.
Thus the probability for a system with initial value $\omega$ to reach
$\mathbf{1}$ consensus equals $\omega$.  As a consequence, a tiny fraction of
very stubborn $\mathbf{1}$ voters (those with flip rates $r\ll 1$) leads to a
probability of reaching $\mathbf{1}$ consensus that is arbitrarily close to
one.

\begin{figure}
\begin{center}
\includegraphics[height=0.29\textwidth,width=0.4\textwidth]{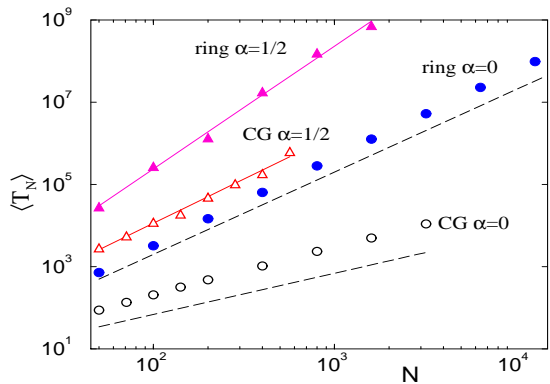}
\caption{Average consensus time $\langle T_N\rangle$ for $10^4$ realizations
  of the HVM on: (a) a complete graph (CG) of $N$ nodes (open symbols), and
  (b) ring of $N$ nodes (filled symbols), with circles and triangles
  corresponding to $\alpha=0$ and $\frac{1}{2}$.  The solid lines are
  power-law data fits, with slopes 2.03 for the CG and 2.98 for the ring,
  compared to the values 2 and 3 from theory (Eq.~\eqref{TN} and immediately
  after Eq.~\eqref{eq:<T> VM Z1 sym}).  The dashed lines are the exact
  results for the homogeneous voter model: (a) $T_N=N\ln 2$ on the CG and (b)
  $T_N\sim N^2$ on the ring.}
\label{T}
\end{center}
\end{figure}

To determine the average consensus time $\langle T_N\rangle$ for a population
of $N$ heterogeneous voters, we focus on the distribution of intrinsic rates
$p(r)=A\,r^{-\alpha}$, with $r\in (0,r_{+}]$ and $\alpha$ in the range
$[0,1)$ so the distribution is normalizable.  For convenience in comparing
cases with different $\alpha$, we fix the average flip rate of the entire
population $\langle r\rangle=1$.  These conditions give $r_{+}=
\frac{2-\alpha}{1-\alpha}$ and $A=(2-\alpha) r_{+}^{\alpha-2}$.  Although the
lower limit of the flip rate distribution is zero, the smallest rate $r_-$
among a finite population of $N$ voters is non-zero and is determined by the
extremal criterion~\cite{extreme}
\begin{equation}
  \int_0^{r_-}\!\! A\,r^{-\alpha}\,dr=N^{-1}\,,
\end{equation}
which gives $r_-\sim N^{-1/(1-\alpha)}$.  As we shall see, these stubbornest
voters control the consensus time.
 
We take the initial condition that each voter is independently in the
$\mathbf{0}$ or the $\mathbf{1}$ state with probability $\frac{1}{2}$.  For
voters on a complete graph of $N\gg 1$ nodes, we now follow the analysis of
the closely-related invasion process on heterogeneous networks \cite{SAR08}.
We partition voters according to their flip rates and denote by $\rho_r$ the
density of $\mathbf{1}$ voters that have flip rate in the range $[r,r+\Delta
r]$.  The evolution of $\rho_r$ is governed by a Fokker-Planck equation whose
drift velocity drives each of the densities $\rho_r$ to the common value
$\rho$ in a convergence time scale that is of the order of ${1}/{r}$.
Subsequently $\rho$ evolves in the same manner as the homogeneous voter model
on the complete graph with an {\em effective} population size $N_{\rm
  eff}=N\langle 1/r\rangle$.  Because the consensus time of the classic voter
model on the complete graph is proportional to this effective size, we
obtain, for the HVM:
\begin{equation}
\label{Neff}
\langle T_N\rangle\sim N\langle 1/r\rangle\,.
\end{equation}

Heterogeneity hinders the approach to consensus because $\langle 1/r\rangle >
1/\langle r\rangle$.  The dependence of Eq.~\eqref{Neff} arises because a
voter with flip rate $r$ effectively corresponds to $1/r$ voters with flip
rate 1.  For the power-law distribution of flip rates $p(r)=Ar^{-\alpha}$,
Eq.~\eqref{Neff}, in conjunction with $r_-\sim N^{-1/(1-\alpha)}$, yields
$\langle 1/r\rangle\sim N^{\alpha/(1-\alpha)}$.  Thus
\begin{equation}
\label{TN}
\langle T_N\rangle \sim
\begin{cases} N\ln N & \alpha=0\,, \\N^{1/(1-\alpha)}&
  0<\alpha<1\,,
\end{cases}
\end{equation}
in agreement with simulation results (Fig.~\ref{T}).  Notice that the
convergence time for the stubbornest voters, $1/r_-\sim N^{1/(1-\alpha)}$, is
of the same order as the consensus time; evidently, this subtlety does not
affect our simulation results.  Finally, if the lower limit of the
distribution of flip rates is strictly greater than zero, then the mean
consensus time is linear in $N$.

In one dimension, the HVM organizes into alternating domains of like-minded
voters at long times, and consensus is reached when all the intervening (and
mobile) domain walls annihilate.  This complete annihilation occurs when a
single domain wall explores on the order of $N$ nodes.  Thus consider the
motion of a single domain wall between nodes $i-1$ and $i$ --- all voters to
the left of $i$ are in state $\mathbf{0}$ and all other voters are in state
$\mathbf{1}$.  In a time interval $dt$, the probabilities that this domain
wall hops one step to the right and to the left are, respectively,
\begin{equation}
\label{pq}
p_i=r_i\Delta t\,,\quad
q_i=r_{i-1}\Delta t\,.
\end{equation}
The crucial point is that hopping probabilities at adjacent nodes are
anti-correlated --- if the bias at node $i$ is to the right (corresponding to
$r_i>r_{i-1}$), then it is more likely that $r_{i+1}<r_i$ and the bias at
node $i+1$ will be to the left.  More precisely, for three consecutive rates
$(r_{i-1},r_i,r_{i+1})$ with the constraint $r_i>r_{i-1}$, their relative
sizes may equiprobably be SML, SLM, or MLS, where S, M, L denotes the
smallest, middle, and largest of these three rates.  The latter two cases
correspond to a leftward bias between nodes $i$ and $i+1$, which thus occurs
with probability 2/3.

With the hopping probabilities $q_i$ and $p_i$, the mean first-passage time
$\tau$ for a particle to travel from $i=0$ to $i=N$ in the finite interval
$[0,N]$ is known~\cite{NG88,L89,MK89}:
\begin{equation}
\label{tau}
\tau=\sum^{N-1}_{k=0}\frac{1}{p_k}+\sum^{N-2}_{k=0}\frac{1}{p_k}
\sum^{N-1}_{i=k+1}\,\,\prod^i_{j=k+1}\frac{q_j}{p_j}~.
\end{equation}
In the Sinai problem~\cite{S82}, where the $p_i$ and $q_i$ are independent,
identically-distributed random variables, $\tau$ grows as $e^N$~\cite{NG88}.
For the HVM, the anti-correlated hopping probabilities \eqref{pq} lead to
substantial cancellations in the above product and yields
\begin{equation}
\tau =N\left<\frac{1}{r}\right>+
 \frac{(N-1)N}{2}\left<\frac{1}{r}\right>~.
\label{eq:<T> VM Z1 sym}
\end{equation}
Using our previous result for $\langle 1/r\rangle$, we thus obtain $\langle
T_N\rangle\sim N^{2+\alpha/(1-\alpha)}$, which agrees well with our numerical
simulations shown in Fig.~\ref{T}(b).  \smallskip

\noindent{\em Partisan Voter Model (PVM):}
Without being pejorative, define state $\mathbf{1}$ as ``democrat'' and state
$\mathbf{0}$ as ``republican''.  In the PVM, each voter has a fixed and
innate preference for democrat or republican.  Equivalently, each voter
experiences its own random field.  A voter can therefore exist in one of four
states: a ``concordant democrat'' is a democratic voter in its preferred
$\mathbf{1}$ state, while ``discordant democrat'' is a democratic voter that
happens to be in the $\mathbf{0}$ state.  Complementary definitions apply for
``concordant republican'' and ``discordant republican''.

Denote the densities of these four types of voters as $D_c$, $D_d$, $R_c$,
and $R_d$, respectively.  The density of voters $\rho$ that happen to be in
the $\mathbf{1}$ state (current democrats) is $\rho=D_c+R_d$.  In a single
update event, a voter in a social network is randomly selected and it selects
a random neighbor.  If these two voters are in the same state, nothing
happens.  If the pair is in different opinion states, the initial voter
changes its state as follows:
\begin{itemize}
\item if the voter becomes aligned with its preference, the change occurs
  at rate $1+\epsilon$;
\item if the voter becomes anti-aligned with its preference, the change
  occurs at rate $1-\epsilon$.
\end{itemize}
Thus $\epsilon$ quantifies the strength of the intrinsic preference, or
partisanship.  If $\epsilon=1$, each voter becomes a zealot that never
changes opinion after aligning with its innate preference, while $\epsilon=0$
recovers the classic voter model.  A similar dichotomous rate arises for
catalysis on a disordered surface~\cite{FKR95}, where surface heterogeneity
controls the adsorption rate of different reactants on the surface.

By analyzing the outcomes from all possible pairs of opposite-opinion voters,
the rate equations for the densities $D_c$ and $D_d$ in the mean-field limit
are:
\begin{align}
\begin{split}
\label{ME}
&\dot D_c= ~~2\epsilon\, D_c D_d 
+(1+\epsilon)\, D_d R_d  - (1-\epsilon)\, D_c R_c\,, \\
&\dot D_d= -2\epsilon\, D_c D_d 
+(1-\epsilon)\, D_c R_c - (1+\epsilon)\, D_d R_d\,. \\
\end{split}
\end{align}
The equations for $R_c$ and $R_d$ are obtained from \eqref{ME} by
interchanging $R\leftrightarrow D$.  Note that $\dot D_c+\dot D_d=\dot
R_c+\dot R_d=0$, which expresses the conservation of voters of any type.

\begin{figure}[ht]
\begin{center}
\includegraphics[height=0.23\textwidth,width=.375\textwidth]{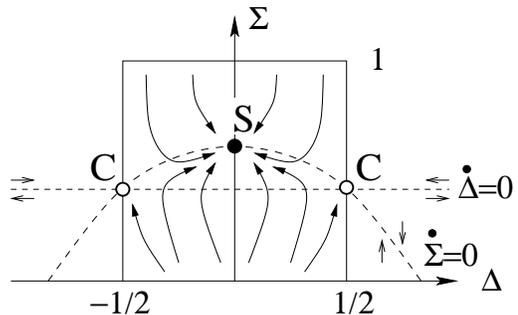}
\caption{Flow diagram (schematic) in the physical portion of the $\Delta$-$\Sigma$
  phase plane (inside square).  The open circles denote the fixed (saddle)
  points that correspond to consensus (C), while the filled dot denotes the
  (stable) self-centered fixed point (S).  The small arrows on either side of
  the nullclines $\dot\Delta=0$ and $\dot \Sigma=0$ (dashed) indicate the local
  flow of $\Delta$ or $\Sigma$.}
\label{FP}
\end{center}
\end{figure}

Let $D$ and $R$ denote the density of intrinsic democrats and republicans,
respectively.  For simplicity, we specialize to the symmetric case of
$D=R=\frac{1}{2}$, so that the density of democrats of any kind, concordant
and discordant, is given by $D_c+D_d=D=\frac{1}{2}$; similarly,
$R_c+R_d=\frac{1}{2}$.  Using these relations,
$\rho=D_c+\frac{1}{2}-R_c\equiv \frac{1}{2}+\Delta$.  In terms of the sum
$\Sigma\equiv D_c+R_c$ and the difference $\Delta\equiv D_c-R_c$ in the densities
of concordant voters, Eqs.~\eqref{ME} simplify to
\begin{align}
\label{DH}
\begin{split}
&\dot \Delta = \epsilon\Delta - 2\epsilon \Sigma\Delta\,,\\
&\dot \Sigma =\tfrac{1}{2}(1+\epsilon) - \Sigma - 2\epsilon \Delta^2\,.
\end{split}
\end{align}

For $\epsilon=0$ (classic voter model), $\dot\Delta=\dot\rho=0$, and the
average density of voters in either opinion state is conserved. Because $\dot
\Sigma =\frac{1}{2}(1-2\Sigma)$, the density of concordant voters is driven to
$\frac{1}{2}$ in the final consensus state.  For general $0<\epsilon<1$,
there are two fixed points (Fig.~\ref{FP}):
\begin{itemize}
\item {\em Self-centered} (S): $\Delta^*=0$ and
  $\Sigma^*=\frac{1}{2}(1+\epsilon)$.  Each voter tends to internal concordance at
  the expense of consensus.
\item {\em Consensus} (C): $\Delta^*=\pm \frac{1}{2}$ and $\Sigma^*= \frac{1}{2}$.
  One half of all the voters are intrinsically concordant.
\end{itemize}

To infer the global flow in the $\Delta$-$\Sigma$ plane, we determine the
nullclines $\dot\Delta=0$ and $\dot \Sigma=0$ (given by $\Sigma=\frac{1}{2}$
and $\Sigma=\frac{1}{2}(1+\epsilon)-2\epsilon\Delta^2$, respectively), and
study the linearized rate equations about each fixed point.  Both eigenvalues
are negative at the self-centered fixed point S, while the eigenvalues have
different signs at the consensus fixed points C.  Thus if voters have innate
preferences, small deviations from the consensus fixed points will grow and
the population will be driven to the S fixed point, where each agent tends to
align with its innate preference.  As the partisanship strength $\epsilon$
increases, each individual is more likely to be aligned with its innate
preference, but with a concomitant lack of consensus.

\begin{figure}[ht]
\begin{center}
\includegraphics[width=.375\textwidth]{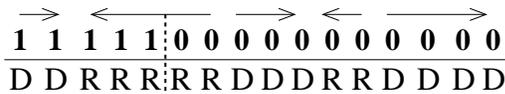}
\caption{State of the PVM in one dimension.  The letters D and R denote the
  intrinsic preference of each voter, while the current state of the voters
  is given the string of $\mathbf{0}$s and $\mathbf{1}$s.  A single domain
  wall is shown by the dashed line and the bias that it experiences is
  indicated by the arrows.  }
\label{PVM}
\end{center}
\end{figure}

\begin{figure}[ht]
\begin{center}
\includegraphics[height=0.3\textwidth,width=0.4\textwidth]{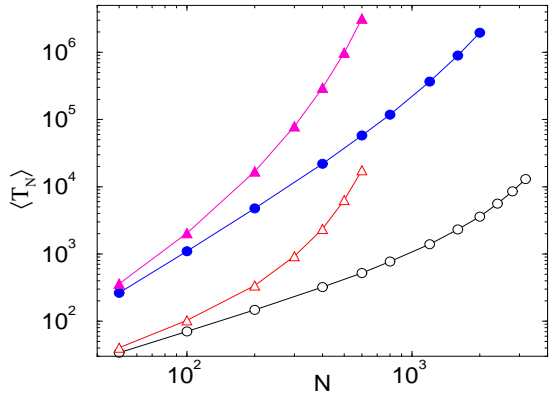}
\caption{Average consensus time $\langle T_N\rangle$ versus number of voters
  $N$ for the PVM on: (a) the complete graph (open symbols) and (b) a ring
  (filled symbols).  The circles and triangles are simulation data for 10000
  realizations with $\epsilon=0.05$ and 0.15, respectively.  The lines are
  guides to the eye.}
\label{fig:asym}
\end{center}
\end{figure}

For a finite system, however, the only true fixed points of the stochastic
dynamics of the PVM are those that correspond to consensus.  Since the flow
in the $\Delta$-$\Sigma$ phase plane is driven away from these fixed points, the
time to reach consensus should scale exponentially in the population size.
We can understand this behavior easily in one dimension because now the
dynamics of single domain walls map exactly to the motion of a particle in a
random potential (the Sinai model~\cite{S82}).  As illustrated in
Fig.~\ref{PVM}, strings of consecutive democrats or republicans give rise to
potential barriers that domain walls have to surmount to annihilate each
other and allow the system to reach consensus.  In a system of length $N$,
the mean time for a domain wall to move a distance $N$ therefore scales as
$e^{N}$~\cite{S82,NG88,L89,MK89}.  Numerical simulations of the PVM on the
complete graph and on the one-dimensional periodic lattice
(Fig.~\ref{fig:asym}) are consistent with this prediction.  We also checked
that qualitatively similar behavior arises when the PVM is generalized to
allow for heterogeneity in the flip rate of each voter.

To summarize, we extended the voter model to incorporate the realistic
features of heterogeneity and partisanship.  Both generalizations are
characterized by a much slower approach to consensus than in the classic
voter model.  When voters are partisan, their individual preferences dominate
over collectivism, and it is only by exponentially rare events that consensus
can ultimately be achieved.  These models offer a step toward the
quantitative modeling of social phenomena, such as threshold models of
collectivism~\cite{G78} and social experiments on incentive-driven consensus
formation~\cite{KJTW}.  Particularly interesting behavior seems to arise when
the two opinion states are inequivalent; in this situation, partisanship for
the unfavorable state may prevent consensus to the favorable state.

\begin{acknowledgments}
  We thank James Fowler for a helpful discussion and literature advice, as
  well as Serge Galam and Gleb Oshanin for relevant references.  NM
  acknowledges financial support by the Grants-in-Aid for Scientific Research
  (Nos.\ 20760258 and 20540382) from MEXT, Japan.  NG was supported by travel
  funds from the Direction g\'en\'erale de l'armement.  SR acknowledges
  support from the US National Science Foundation grant DMR0906504.
\end{acknowledgments}


\begin{thebibliography}{99}

\bibitem{L85} T.~M.~Liggett, {\it Interacting Particle Systems},
  (Springer-Verlag, New York, 1985).

\bibitem{K02} P. L. Krapivsky, Phys.\ Rev.\ A {\bf 45}, 1067 (1992).

\bibitem{spin} J.~D.~Gunton, M.~San~Miguel, and P.~S.~Sahni, in: {\it Phase
    Transitions and Critical Phenomena}, Vol.~8, eds.\ C.~Domb and
  J.~L.~Lebowitz (Academic, NY 1983); A.~J.~Bray, Adv.\ Phys.\ {\bf 43}, 357
  (1994).

\bibitem{CFL} See {\it e.g.}, C. Castellano, S. Fortunato, and V. Loreto,
  Rev.\ Mod.\ Phys.\ {\bf 81}, 591 (2009).

\bibitem{BFK96} E. Ben-Naim, L. Frachebourg, and P. L. Krapivsky, Phys.\
  Rev.\ E {\bf 53}, 3078 (1996)
        
\bibitem{DCCH01} I. Dornic, H. Chat\'e, J. Chave, and H. Hinrichsen, Phys.\
  Rev.\ Lett.\ {\bf 87}, 045701 (2001).

\bibitem{SR05} V. Sood, S. Redner, Phys.\ Rev.\ Lett.\ {\bf 94}, 178701
  (2005)

\bibitem{SAR08} V. Sood, T. Antal, and S. Redner, Phys.\ Rev.\ E {\bf 77},
  041121 (2008).

\bibitem{FDC09} J. H. Fowler, C. T. Dawes, and N. A. Christakis, Proc.\
  Natl.\ Acad.\ Sci.\ (USA) {\bf 106}, 1720 (2009); J. E. Settle,
  C. T. Dawes, and J. H. Fowler, Polit.\ Res.\ Quarterly {\bf 62}, 601 (2009).

\bibitem{G78} M. Granovetter, Am.\ J.  Sociol.\ {\bf 83}, 1420 (1978).

\bibitem{M03} M. Mobilia, Phys.\ Rev.\ Lett.\ {\bf 91}, 028701 (2003);
  M. Mobilia, A. Petersen, and S. Redner, J. Stat.\ Mech.\ P08029, (2007).

\bibitem{GF07} S. Galam and F. Jacobs, Physica A {\bf 381}, 366 (2007).

\bibitem{G97} S. Galam,  Physica A {\bf 238}, 66 (1997).

\bibitem{KJTW} M. Kearns, S. Judd, J. Tan, and J. Wortman, Proc.\ Natl.\
  Acad.\ Sci.\ (USA) {\bf 106}, 1347 (2009).

\bibitem{extreme} J. Galambos, {\it The Asymptotic Theory of Extreme Order
    Statistics} (Krieger Publishing Co., Malabar, FL, 1987).

\bibitem{NG88} S. H. Noskowicz and I. Goldhirsch, Phys.\ Rev.\ Lett.\ {\bf
    61}, 500 (1988).

\bibitem{L89} P. Le Doussal, Phys.\ Rev.\ Lett.\ {\bf 62}, 3097
  (1989).

\bibitem{MK89} K. P. N. Murthy and K. W. Kehr, Phys.\ Rev.\ A {\bf
    40}, 2082 (1989).

\bibitem{S82} Ya.\ G. Sinai, Theor.\ Probab.\ Appl.\ {\bf 27} 256 (1982).

\bibitem{FKR95} L. Frachebourg, P. L. Krapivsky, and S. Redner, Phys.\ Rev.\ Lett.\
  {\bf 75}, 2891 (1995).

\end{thebibliography}
\end{document}